\documentclass[%
 aip,
 amsmath,amssymb,
preprint,%
]{revtex4-1}

\usepackage{graphicx}
\usepackage{dcolumn}
\usepackage{bm}

\usepackage[utf8]{inputenc}
\usepackage[T1]{fontenc}
\usepackage{mathptmx}
\usepackage{etoolbox}

\makeatletter
\def\@email#1#2{%
 \endgroup
 \patchcmd{\titleblock@produce}
  {\frontmatter@RRAPformat}
  {\frontmatter@RRAPformat{\produce@RRAP{*#1\href{mailto:#2}{#2}}}\frontmatter@RRAPformat}
  {}{}
}%
\makeatother
\begin{document}

\preprint{AIP/123-QED}

\title{Cross-beam energy transfer in conditions relevant to direct-drive implosions on OMEGA}
\author{K. L. Nguyen}
\email{lnguy17@ur.rochester.edu,  jpal@lle.rochester.edu}
\affiliation{
\mbox{Laboratory for Laser Energetics, University of Rochester, Rochester, New York, 14623, USA}
}
\affiliation{
\mbox{Department of Physics \& Astronomy, University of Rochester, Rochester, New York, 14623, USA}
}
\affiliation{
Los Alamos National Laboratory, Los Alamos, New Mexico, 87545, USA
}
\author{L. Yin}
\affiliation{
Los Alamos National Laboratory, Los Alamos, New Mexico, 87545, USA
}
\author{B. J. Albright}
\affiliation{
Los Alamos National Laboratory, Los Alamos, New Mexico, 87545, USA
}
\author{D. H. Edgell}
\affiliation{
\mbox{Laboratory for Laser Energetics, University of Rochester, Rochester, New York, 14623, USA}
}
\author{R. K. Follett}
\affiliation{
\mbox{Laboratory for Laser Energetics, University of Rochester, Rochester, New York, 14623, USA}
}
\author{D.Turnbull}
\affiliation{
\mbox{Laboratory for Laser Energetics, University of Rochester, Rochester, New York, 14623, USA}
}
\author{D. H. Froula}
\affiliation{
\mbox{Laboratory for Laser Energetics, University of Rochester, Rochester, New York, 14623, USA}
}
\author{J. P. Palastro}
\affiliation{
\mbox{Laboratory for Laser Energetics, University of Rochester, Rochester, New York, 14623, USA}
}

\date{\today}

\begin{abstract}
In cross-beam energy transfer (CBET), the interference of two laser beams ponderomotively drives an ion-acoustic wave that coherently scatters light from one beam into the other. This redirection of laser beam energy can severely inhibit the performance of direct-drive inertial confinement fusion (ICF) implosions. To assess the role of nonlinear and kinetic processes in direct-drive-relevant CBET, the energy transfer between two laser beams in the plasma conditions of an ICF implosion at the OMEGA laser facility was modeled using particle-in-cell simulations. For typical laser beam intensities, the simulations are in excellent agreement with linear kinetic theory, indicating that nonlinear processes do not play a role in direct-drive implosions. At higher intensities, CBET can be modified by pump depletion, backward stimulated Raman scattering, or ion trapping, depending on the plasma density. 

\end{abstract}

\maketitle


\section{\label{sec:level1}INTRODUCTION}

In direct-drive inertial confinement fusion (ICF), an ensemble of laser beams directly illuminate a spherical target containing deuterium-tritium fuel.\cite{CraxtonReview2015} Initially, the laser beams ionize and heat the target surface, forming an ablation front surrounded by a lower-density plasma corona. Thereafter, the laser beam energy is absorbed in the plasma corona and conducted to the ablation front by electron thermal transport. Continued ablation of the target accelerates the fuel inward until it converges at the target center. If the converged fuel has sufficient thermal energy or can be confined long enough, it will ignite and undergo thermonuclear burn. Compared to other ICF schemes, direct-drive has the potential to couple a larger fraction of the laser energy to the target, but like other schemes, it is susceptible to laser-plasma instabilities that can inhibit implosion performance.


Among these instabilities, cross-beam energy transfer (CBET) has been identified as the leading cause of decreased energy coupling in direct-drive implosions. \cite{RandallCBET, KruerCBET, DanaAPSDPP2008,froula2012increasing} In direct-drive-relevant CBET, two frequency-degenerate laser beams ponderomotively excite an ion acoustic wave that scatters energy from one beam into the other. The outward flow of the plasma corona enhances the excitation by shifting the ion acoustic wave frequency into resonance. For a typical implosion at the OMEGA laser facility, CBET transfers energy from the center of an incoming beam to the periphery of other beams, which reduces laser absorption by 10\% to 20\%. \cite{igumenshchev2010crossed,igumenshchev2012crossed} The  redirection of laser light also degrades absorption uniformity, which can contribute significantly to implosion asymmetries. \cite{EdgellCBETnonuniform,Arnaudasymmetries}




The impact of reduced absorption and symmetry on implosion performance has motivated the development of reduced CBET models that are suitable for implementation into radiation hydrodynamics simulations. \cite{FollettRayCaustic, ArnaudRay1, ArnaudRay2, DebayleRay} To maintain a computational cost that is commensurate with radiation hydrodynamics codes, the models combine ray tracing with steady-state coupled-mode equations instead of using more fundamental wave or particle-based approaches.\cite{MyattLPSE,bowers2008ultrahigh} As a result, these models neglect a number of nonlinear processes that may affect, or even saturate, CBET, including ion trapping, two-ion decay, nonlinear sound waves, and stochastic heating. \cite{williams2004effects,niemann2004observation,michel2012stochastic,michel2013saturation,albright2016multi,chapman2017longitudinal,yin2019saturation,huller2020crossed,nguyen2021cross,seaton2022cross1} 

While several experiments have observed CBET saturation,\cite{KirkwoodSat, TurnbullSat,hansen2021cross,hansen2022cross} recent experiments on the OMEGA laser were the first to identify a saturation mechanism.\cite{hansen2021cross} Specifically, the OMEGA experiments showed that CBET can saturate through a resonance detuning caused by trapping-induced modifications to the ion distribution functions \cite{hansen2021cross,nguyen2021cross,hansen2022cross}. The experiments employed high-intensity laser beams ($\sim$$10^{15}$ $\mathrm{W/cm}^2$) crossing in a low-density gas-jet plasma ($n_e$ $\leq$ 1.2\% $n_c$, where $n_e$ is the electron density, $n_{c} = \epsilon_0 m_e\omega^2/e^2 $ and $\omega$ is the laser frequency) with electron and ion temperatures of only a few hundreds of eV. In contrast, CBET in direct-drive ICF occurs between lower intensity laser beams ($\sim$$10^{14}$ $\mathrm{W/cm}^2$) in a much hotter and denser plasma. This raises the question: what, if any, saturation mechanisms occur in direct-drive-relevant conditions?

This manuscript presents the results of particle-in-cell (PIC) simulations which indicate that nonlinear or saturation processes do not play a role in direct-drive-relevant CBET. The simulations modeled the energy transfer between two laser beams using plasma conditions and intensities extracted from a radiation hydrodynamics simulation of an OMEGA implosion. For direct-drive-relevant intensities, the PIC simulations are in excellent agreement with the linear, steady-state, kinetic theory of CBET. For the same plasma conditions but higher intensities, different nonlinear processes are observed to affect CBET depending on the plasma density: At $n_e/n_c = 30\%$, CBET can saturate due to pump depletion; at $n_e/n_c = 20\%$, the seed beam can become unstable to backward stimulated Raman scattering, which reduces the apparent energy transfer; and at $n_e/n_c = 10\%$, ion trapping can enhance CBET by reducing Landau damping of the driven ion acoustic wave.

The remainder of this manuscript is organized as follows: Section II describes a commonly used reduced model for CBET in direct-drive. Section III presents the setup for the PIC simulations. Section IV demonstrates the agreement between CBET and linear theory for direct-drive-relevant intensities. Section V discusses the nonlinear processes that can affect CBET at high intensities. Section VI concludes the manuscript with a summary of the results.

\begin{figure*}
\includegraphics[scale=.9]{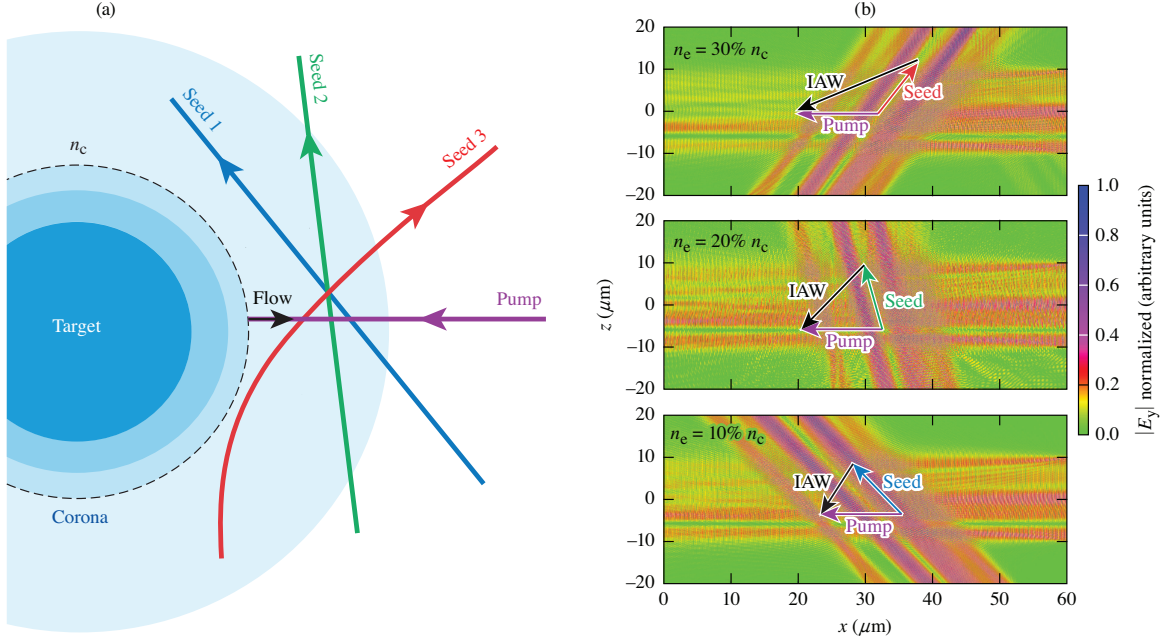}
\caption{(a) A schematic of CBET in a typical direct drive implosion. An incoming ray (purple), anti-parallel to the hydrodynamic flow, interacts with rays from the edges of other beams (blue, green, and red). (b) Total electromagnetic field amplitudes of the pump and seed, illustrating the simulated interaction geometries at 30\%, 20\%, and 10\% $n_c$.}
\label{fig:f1}
\end{figure*}

\section{\label{sec:level1}REDUCED MODEL FOR CBET IN DIRECT DRIVE}

In a direct-drive implosion, CBET occurs between beams with the same frequency and tends to redistribute energy from the central portion of an incoming laser beam to the periphery of another laser beam. Figure ~\ref{fig:f1}(a) illustrates the intersection of rays within different beams in the expanding plasma surrounding the target. The incident ray (purple) travels towards the target and intersects rays at the edges of other beams (blue, green, and red).

The ponderomotive force of each crossing pair drives an ion acoustic wave with a frequency ($\Omega$) and wavevector (\textbf{k}) determined by the conditions: $\Omega = \omega_0 - \omega_1$ and $\mathbf{k} = \mathbf{k}_0 - \mathbf{k}_1$, where the subscripts 0 and 1 denote two crossing rays. The density perturbation of the ion acoustic wave acts like a grating, coherently scattering light from one ray into the other. 

A commonly used reduced model for CBET approximates rays within each beam as plane waves and assumes a linear, steady-state plasma response. In this model, the intensities of two intersecting rays evolve according to
\begin{eqnarray}\label{eq:CME}
(\mathbf{v}_a\cdot\nabla)I_a = g_{ab}I_bI_a
\end{eqnarray}
where a $\neq$ b can take the values 0 or 1, $I_{a}$ is the intensity, $\mathbf{v}_a = (1-\omega_{p}^2/\omega_a^2)^{1/2}c\mathbf{k}_a/k_a$ is the group velocity, $\omega_p = (e^2n_e/\epsilon_0m_e)^{1/2}$ is the plasma frequency, and 
\begin{eqnarray}
g_{ab} = -\frac{e^2k^2\eta_{ab}}{2\epsilon_{0}m_{e}^2\mathrm{v}_b\omega_{b}^2\omega_a}\mathrm{Im}\left[\Gamma(\Omega,\mathbf{k})\right]
\end{eqnarray}
with $\eta_{01} = -\eta_{10} = 1$. The kinetic coupling factor, 
\begin{eqnarray}\label{eq:KCF}
\Gamma(\Omega,\mathbf{k})=\frac{\chi_e(\Omega,\mathbf{k})[1+\chi_i(\Omega,\mathbf{k})]}{1+\chi_i(\Omega,\mathbf{k})+\chi_e(\Omega,\mathbf{k})},
\end{eqnarray}
captures the response of the plasma to the ponderomotive drive. Here $\chi_e$ and $\chi_i$ are the electron and ion susceptibilities:
\begin{eqnarray}
\chi_e(\Omega,\mathbf{k}) = \frac{e^2}{\epsilon_0m_ek^{2}}\int \frac{\mathbf{k}\cdot \nabla_{\mathbf{v}} f_{e}}{\Omega-\mathbf{k}\cdot\mathbf{u}_f-\mathbf{k}\cdot\mathbf{v}}d\mathbf{v}
\end{eqnarray}
\begin{eqnarray}
\chi_i(\Omega,\mathbf{k}) =\sum_{s} \frac{Z_s^2e^2}{\epsilon_0m_sk^{2}}\int\frac{\mathbf{k}\cdot \nabla_{\mathbf{v}} f_{s}}{\Omega-\mathbf{k}\cdot\mathbf{u}_f-\mathbf{k}\cdot\mathbf{v}}d\mathbf{v},
\end{eqnarray}
$f_e$ and $f_s$ are the electron and ion distribution functions in the stationary frame of the plasma, $Z_s$ is the ion charge, $m_s$ is the ion mass, and $\textbf{u}_f$ is the hydrodynamic flow velocity. 

The maximum rate of energy transfer between two rays occurs when the ion acoustic wave is driven on resonance, i.e., when
\begin{eqnarray}\label{eq:resonance1}
\Omega = \Omega_A + \mathbf{k}\cdot\mathbf{u}_f,
\end{eqnarray}
where $\Omega_A$ is the real, natural mode frequency of the ion acoustic wave in the absence of flow. The resonant drive frequency minimizes the absolute value of the dielectric constant $|\epsilon(\Omega,\mathbf{k})| = |1 + \chi_e(\Omega,\mathbf{k}) + \chi_i(\Omega,\mathbf{k})|$, which maximizes Im[$\Gamma(\Omega,\mathbf{k})$] and $g_{ab}$. 

The flow velocity shifts the frequency of the ion acoustic wave, allowing for resonant excitation even when each beam (or ray) has the same frequency. Specifically, 
when $\omega_0 - \omega_1 = \Omega = 0$, the resonance condition can be satisfied if
\begin{eqnarray}\label{eq:resonance2}
\mathbf{k}\cdot\mathbf{u}_f = -\Omega_A.
\end{eqnarray}
As depicted in Fig. \ref{fig:f1}(a), the flow velocity is directed radially outward. Thus, in order to satisfy Eq. \ref{eq:resonance2}, the wavevector of the ion acoustic wave must point towards the target. Equivalently, ray 0 must have a smaller local angle of incidence (i.e., be more anti-parallel to the flow) than ray 1: $\mathbf{k}_0\cdot\mathbf{u}_f <  \mathbf{k}_1\cdot\mathbf{u}_f$.

To determine the direction of energy transfer, it is convenient to perform a Galilean transformation into the stationary frame of the plasma. In this frame, denoted by a prime $'$, the frequencies of the rays are given by $\omega'_a = \omega_a - \mathbf{k}_a\cdot\mathbf{u}_f$.  Thus, ray 0 is blue shifted relative to ray 1, such that $\omega'_0 = \omega'_1 + \Omega_A$. This indicates that ray 0 transfers energy to ray 1 and explains the redistribution of energy from the center of incoming beams, where rays have smaller angles of incidence, to the periphery of other beams, where rays have larger angles of incidence. To indicate the direction of energy transfer, 0 will denote the pump and 1 the seed for the remainder of the manuscript. 

In all of the calculations presented here, the pump will have the minimum (zero) angle of incidence and propagate anti-parallel to the flow [see Fig. \ref{fig:f1}(b) for the geometries]. Upon using $\mathbf{k}_0\cdot\mathbf{u}_f = -k_0u_f$ in Eq. \ref{eq:resonance2}, one can rewrite the resonance condition for frequency-degenerate beams as  
\begin{eqnarray}\label{eq:phase}
\mathrm{v}'_p = u_f\mathrm{sin}(\theta/2),
\end{eqnarray}
where $\textrm{v}'_p$ = $\Omega_A/k$  is magnitude of the ion acoustic wave phase velocity in the stationary frame of the plasma and $\theta$ is the crossing angle between the pump and the seed. Equation \ref{eq:phase} demonstrates that resonance is impossible when $u_f < \mathrm{v}'_p$, regardless of crossing angle.

When the pump and seed propagate in the same direction ($\theta = 0$), Eq.\ref{eq:CME} has the analytical solution\cite{tang1966saturation}
\begin{eqnarray}\label{eq:tang}
I^{\mathrm{out}}_1 = \frac{(1+\beta)\mathrm{exp}[G_{0}(1+\beta)]}{1+\beta \mathrm{exp}[G_{0}(1+\beta)]}I^{\mathrm{in}}_1,
\end{eqnarray}
where the superscripts ``in'' and ``out'' denote the input and output values of intensity and $\beta = I^{\mathrm{in}}_1/I^{\mathrm{in}}_0$. The small signal gain $G_0$ is given by
\begin{eqnarray}\label{eq:G0}
G_0 = k_0L\mathrm{sin}^2(\theta/2)\mathrm{Im}[\Gamma(\Omega,\mathbf{k})]|\mathbf{a}_0^{\mathrm{in}}|^2,
\end{eqnarray}
where $L$ is the interaction length and $|\textbf{a}_0^{\mathrm{in}}|=(e/m_ec\omega_0)(2I_0^{\mathrm{in}}/\epsilon_0\mathrm{v_0})^{1/2}$ is the normalized amplitude of the pump. Using Eq. \ref{eq:tang}, one can calculate the gain including the effect of pump depletion: 
\begin{eqnarray}\label{eq:GD}
G_D = \mathrm{ln}(I^{\mathrm{out}}_1/I^{\mathrm{in}}_1).
\end{eqnarray}
In the limit of small $I^{\mathrm{in}}_1$, $G_D \to G_0$, reproducing the familiar small signal gain result, $I^{\mathrm{out}}_1 = I^{\mathrm{in}}_1e^{G_0}$. 

Oblique intersection of the pump and the seed can be accounted for by replacing the interaction length in Eqs. \ref{eq:tang} and \ref{eq:G0} by $L = d/\sin\theta$, where $d$ is the width of each beam. This approximation was compared to more sophisticated 2D models\cite{mckinstrie1996two,marion2016modeling} of planar-like beams and was found to be in excellent agreement. The approximation also works well for speckled beams when the average intensities are used for $I^{\mathrm{in}}_0$ and $I^{\mathrm{in}}_1$, so long as the gain is moderate ($G_0 \lesssim 1$) and accumulates over multiple speckle widths, which is typically the case in direct-drive CBET. \cite{follett2017full} 

\begin{figure}
\includegraphics[scale=1]{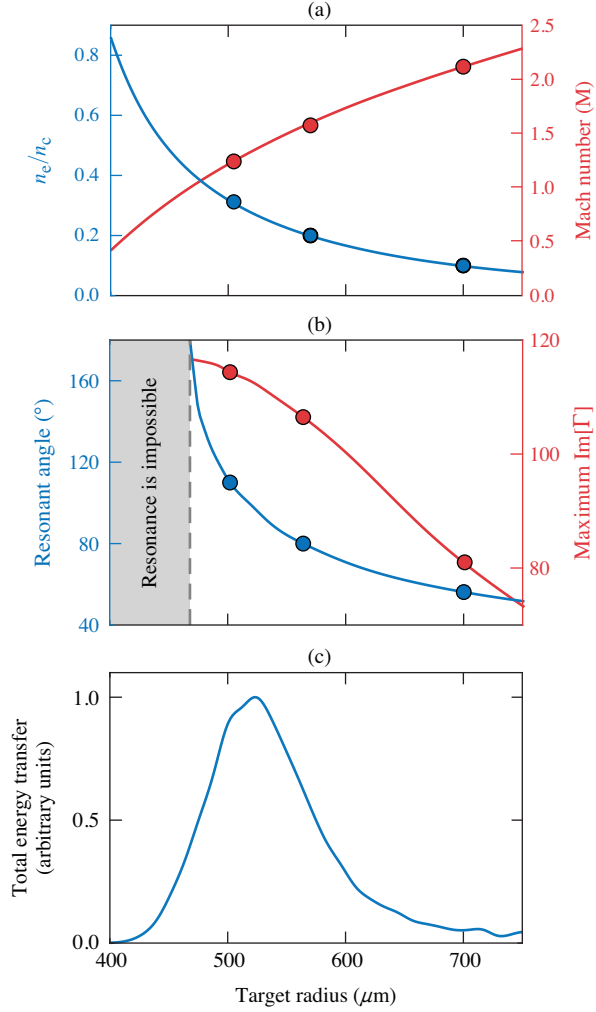}
\caption{(a) Density (blue, left) and flow (red, right) profiles from a LILAC simulation of a direct-drive implosion on OMEGA at peak power. (b) The resonant crossing angle between the pump and seed lasers (blue) maximizes the value of Im[$\Gamma$] (red), and thus the rate of intensity transfer, at each location. (c) The total energy transfer calculated using a 3D ray-based model of CBET, indicating that CBET is most active near 30\% $n_c$. The VPIC simulations were initialized with the plasma conditions at the location of the dots in (a) and (b).}
\label{fig:f2}
\end{figure}
 
\section{\label{sec:level1}SIMULATION SETUP}

The reduced model described above neglects several effects that can modify CBET, including non-steady-state, nonlinear, and collisional processes. In principle, one would like the simplest possible model of CBET that describes experimental observations. In practice, integrated experiments, such as direct-drive implosions, complicate the interpretation of measurements, because several competing processes can lead to similar features in the data. This makes it difficult to isolate and rule out effects. Alternatively, one can use a more sophisticated model to test the validity of the reduced model and the impact of the neglected effects.

The approach here is to test the reduced CBET model against PIC simulations initialized with plasma conditions extracted from a radiation hydrodynamics simulation of an OMEGA implosion. The PIC simulations capture the non-steady-state, nonlinear, and collisional processes that are not included in the reduced model. As a result, the impact of these effects can be assessed by comparing the PIC results to the reduced model. 

The implosion of a warm CH target (shot 96299) was simulated using the 1D radiation hydrodynamics code LILAC. \cite{boehly2011multiple} At peak power, the coronal plasma consists of electrons and two fully ionized ion species: approximately $50\%$ each of hydrogen and carbon. The electron and ion temperatures, $T_e = 2.5\; \mathrm{keV}$ and $T_i = 1.2\; \mathrm{keV}$, are nearly uniform. Figure \ref{fig:f2}(a) displays the electron density and flow profiles. Here the Mach number $\mathrm{M} \equiv u_f/\mathrm{v}'_p$, where $\mathrm{v}'_p \approx 3.6\times10^5 \,\mathrm{m/s}$ is calculated by solving for the root of $\epsilon(\Omega,\mathbf{k}) = 0$ using Maxwellian electron and ion distribution functions. 

In accordance with the flow profile and Eq. \ref{eq:phase}, the crossing angle $\theta$ required for resonant excitation of the ion acoustic wave increases towards the target [Fig. \ref{fig:f2}(b)]. At each location, the resonant angle maximizes the coupling factor $\mathrm{Im}[\Gamma]$. The value of $\mathrm{Im}[\Gamma]$ increases with density towards the target until the flow velocity drops below $\mathrm{M} = 1$ [right axis of Figs. \ref{fig:f2}(a) and (b)]. Past this point, $\mathrm{M<1}$ and resonant excitation is impossible. Figure \ref{fig:f2}(c) demonstrates that CBET is the most energetically significant near $30\% \;n_c$, which is consistent with the maximum of $\mathrm{Im}[\Gamma]$ in Fig. \ref{fig:f2}(b). The total energy transfer at each location was calculated using a 3D ray-based, reduced model of CBET. \cite{edgell2017mitigation} Note that CBET still occurs in the subsonic region (i.e., where $\mathrm{M}<1$), albeit non-resonantly.

The energy transfer between crossing beams was simulated using the Vector Particle-in-Cell code\cite{bowers2008ultrahigh} (VPIC). Three resonant crossing angles were considered: $\theta = 110^\textrm{o}$ at $n_e = 30\% \,n_c$, $\theta = 80^\textrm{o}$ at $n_e = 20\% \,n_c$, and $\theta = 60^\textrm{o}$ at $n_e = 10\% \,n_c$ (see Figs. \ref{fig:f1} and \ref{fig:f2} for geometries and locations). In all cases, the beams were polarized out of the plane of the simulation, i.e., in the $\hat{\mathbf{y}}$-direction; had a width of $d = 20 \,\mu \mathrm{m}$; and were composed of a random distribution of $f/6.7$ speckles.\cite{yin2012trapping} The beam width, while much smaller than that of an actual OMEGA beam ($\sim 800 \, \mu \mathrm{m}$),\cite{lees2021experimentally} results in gains that are large enough to test the salient physics.

The plasma conditions in the VPIC simulations were determined by the LILAC profiles. For each crossing angle, the density profile was modeled as a linear ramp increasing towards the target in the $\hat{\mathbf{x}}$-direction. The ramp was centered at the respective density for each crossing angle, $n_e = 30\%$, $20\%$, and $10\% \,n_c$, and had a  scale length $L_n$ equal to that of the LILAC profile at that density. The initial electron and ion temperatures, $T_e = 2.5\; \mathrm{keV}$ and $T_i = 1.2\; \mathrm{keV}$, were uniform. 

The flow velocity at each density was emulated by redshifting the seed beam relative to the pump. Specifically, the pump beam had a vacuum wavelength $\lambda_0 = 351 \,\mathrm{nm}$ in all cases, while the vacuum wavelength of the seed beam was specified using
\begin{eqnarray}\label{eq:lambdaseed}
\lambda_1 = \lambda_0 + \frac{\lambda_0^2}{2\pi c}ku_f\mathrm{sin}(\theta/2) = \lambda_0 + \frac{\lambda_0^2}{2\pi c}\Omega_A.
\end{eqnarray}
This is equivalent to simulating the interaction in the local rest frame of the plasma. The flow gradient was not included in the VPIC simulations  (see Appendix A for further discussion). The physical parameters are summarized in Table I. 

All of the VPIC simulations included ion-ion, electron-electron, and electron-ion collisions. The binary collision model implemented in VPIC recovers the Landau form for inter-particle collisions in weakly coupled plasmas.\cite{takizuka1977binary,yin2016plasma} The domain of each simulation was $60 \,\mu \mathrm{m} \times 40\, \mu \mathrm{m}$ with cell sizes $\Delta x \times \Delta z \approx 1.0\lambda_D \times 1.0\lambda_D$, where $\lambda_D$ = $(\epsilon_0k_BT_e/n_ee^2)^{1/2}$ is the Debye length, and 512 particles per cell on average. Absorbing boundary conditions were used for the fields, and refluxing boundary conditions were used for the particles: Every particle that left the simulation domain was replaced by a particle injected with a velocity randomly sampled from a fixed-temperature Maxwellian distribution.

\begin{table}
\caption{VPIC simulation parameters determined by a LILAC simulation of an OMEGA implosion. The wavelength shift of the seed $\Delta\lambda = \lambda_1 - \lambda_0$ emulates the local flow velocity in the corona. The electron and ion temperatures, $T_e = 2.5\; \mathrm{keV}$ and $T_i = 1.2 \; \mathrm{keV}$, were uniform in all cases.}
\begin{ruledtabular}
\begin{tabular}{lccc}
$ $&10\% $n_c$&20\% $n_c$&30\% $n_c$\\
\hline
$L_n$ $(\mu \mathrm{m})$ & $300$ & $225$ & $200$ \\
$\theta$ $(\textrm{degrees}) $ & $60$ & $80$ & $110$\\
$\Delta\lambda$ (\AA) & $4.0$ & $4.9$ & $5.9$\\
\end{tabular}
\end{ruledtabular}
\label{tab:t1}
\end{table}

\section{\label{sec:OMI}OMEGA-RELEVANT INTENSITIES}

\begin{figure}
\includegraphics[scale=1]{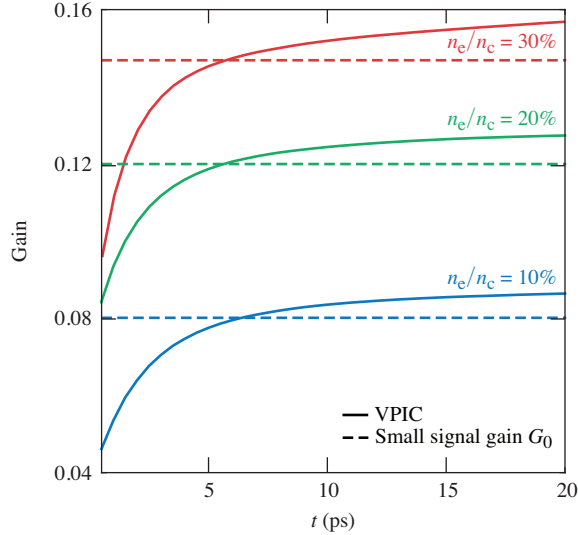}
\caption{Time evolution of the CBET gain for OMEGA-relevant intensities (Table II). At each density, the gain evolves through a transient stage before reaching a near-steady-state value slightly above the small signal gain $G_0$ (Eq. \ref{eq:G0}).}
\label{fig:f3}
\end{figure}

For intensities relevant to a direct-drive implosion on OMEGA, the VPIC simulations are in excellent agreement with the linear, steady-state CBET theory. This is illustrated by Fig. \ref{fig:f3}, which compares $G_0$ (Eq. \ref{eq:G0}) to the gain calculated from VPIC: $G_{\mathrm{V}} = \mathrm{ln}(P_1^{\mathrm{out}}/P_1^{\mathrm{in}})$, where $P_1^{\mathrm{in}}$ and $P_1^{\mathrm{out}}$ are the input and output seed powers. The VPIC gain evolves through a transient stage\cite{divol2019analytical} over the first $\sim$ 5 ps before approaching a near-steady state value. The modest discrepancy between the VPIC gain and $G_0$ results from the relatively small number of speckles in the pump and seed beams.

The pump and seed beams at each density were initialized with average intensities determined by the 3D ray-based model of CBET \cite{edgell2017mitigation} discussed above (see Table II). Inverse bremsstrahlung heating and CBET lower the average intensity of the pump as it propagates from $10\% \, n_c$ to $30\% \, n_c$. Despite the lower pump intensity, the highest gain occurs at $30\% \, n_c$ due to the density scaling of the coupling factor $\mathrm{Im}[\Gamma]$ [Fig. \ref{fig:f2}(b) and Fig. \ref{fig:f3}]. The seed at 30\% $n_c$ corresponds to a ray that originates closers to the center of an OMEGA beam where the intensity is higher, while the seed at 10\% $n_c$ corresponds to a ray further into the periphery of an OMEGA beam where the intensity is lower.

\begin{table}
\caption{Average intensities of the pump and seed at each plasma density for the OMEGA-relevant simulations.}
\begin{ruledtabular}
\begin{tabular}{lccc}
$ $&10\% $n_c$&20\% $n_c$&30\% $n_c$\\
\hline
$I_0^{\mathrm{in}}$ $(\mathrm{W/cm}^2)$ & $9.8\times10^{13}$ & $8.1\times10^{13}$ & $5.5\times10^{13}$ \\
$I_1^{\mathrm{in}}$ $(\mathrm{W/cm}^2)$ & $2.6\times10^{11}$ & $2.5\times10^{12}$ & $1.0\times10^{13}$\\
\end{tabular}
\end{ruledtabular}
\label{tab:t2}
\end{table}

\section{\label{sec:HI}NONLINEAR EFFECTS AT HIGHER INTENSITIES}

The intensities used in Sec. \ref{sec:OMI} are typical of conventional ``hot-spot'' ignition designs for direct-drive ICF. When hydrodynamically scaling from OMEGA to larger, more ignition-relevant scales, the laser pulse power and target surface area increase in a proportion that keeps the incident intensity fixed. There are, however, a number of direct-drive concepts that propose using higher intensities, such as ``fast'' and ``shock'' ignition.\cite{TabakFastIgnition,Clark_2007, BettiShockIgnition} 

This section presents the results of VPIC simulations of CBET between higher intensity laser beams in OMEGA-relevant plasma conditions. For consistency with the low-intensity simulations, the plasma conditions and beam crossing angles were kept the same. The pump intensity was set to $I_0^{\mathrm{in}} = 2\times10^{14} \;\mathrm{W/cm^2}$ in all cases, while the average seed intensity was set to $I_1^{\mathrm{in}} = 1$ or $2\times10^{14} \;\mathrm{W/cm^2}$. These values bracket the threshold for the observed nonlinear phenomena. 

At the higher seed intensity, the simulations show that different nonlinear processes can affect CBET, depending on the density. Specifically, pump depletion is observed at 30\% $n_c$, backward stimulated Raman scattering (BSRS) of the seed at 20\% $n_c$, and ion trapping at 10\% $n_c$. These different processes occur despite the prediction of linear theory that the relative amplitude of the ion acoustic waves should be nearly the same: Calculating 
\begin{eqnarray}
\frac{\delta n_e}{n_{e}}=\frac{c^2k^2}{2\omega_{p}^2} | \Gamma(\Omega,\mathbf{k}) (\mathbf{a}_0^{\mathrm{in}} \cdot \mathbf{a}_1^{\mathrm{in}})|
\end{eqnarray}
at each density provides $\delta n_e/n_e \approx 0.32 \%$ for $I_0^{\mathrm{in}} = I_1^{\mathrm{in}} = 2\times10^{14} \,\mathrm{W/cm^2}$. To reflect the energetic significance of CBET at each location [Fig.\ref{fig:f2}(c)], the simulations will be discussed in order of  descending density (i.e., increasing target radius). 

\begin{figure}
\includegraphics[scale=1]{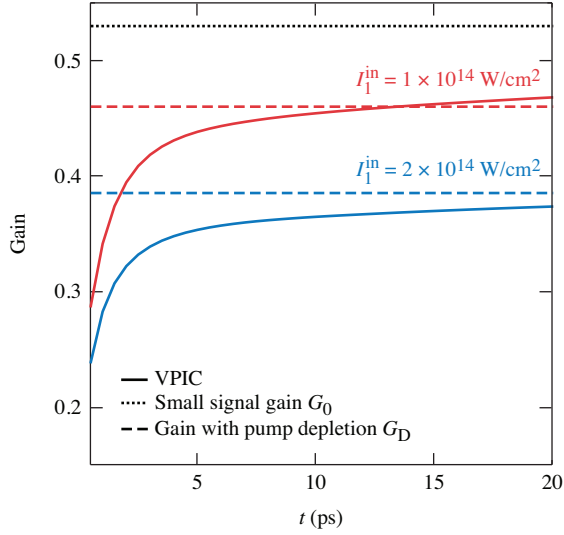}
\caption{Time evolution of the CBET gain for the high pump and seed intensities at 30\% $n_c$. Here, $I_0^{\mathrm{in}} = 2\times10^{14} \,\mathrm{W/cm^2}$. Pump depletion causes the gain to asymptote to a value lower than the small signal gain $G_0$. Consistent with Eq. \ref{eq:tang}, the reduction in gain is larger for the higher seed intensity.}
\label{fig:f4}
\end{figure}

\begin{figure}
\includegraphics[scale=1]{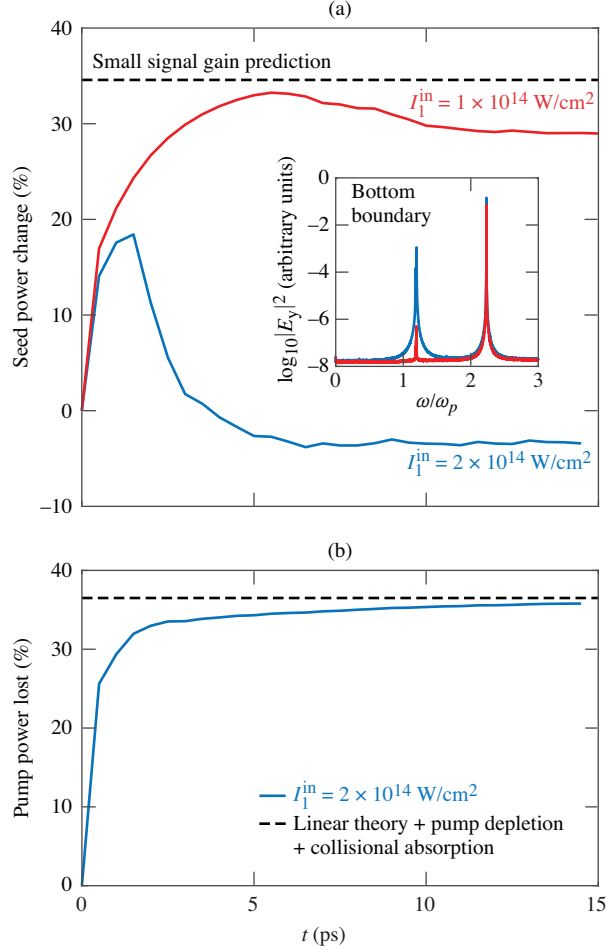}
\caption{\label{fig:wide} (a) Change in the transmitted seed power, $P_1^{\mathrm{out}}/P_1^{\mathrm{in}}-1$, as a function of time for the high pump and seed intensities at 20\% $n_c$. When $I_1^{\mathrm{in}}$ = $1\times10^{14} \mathrm{W/cm^2}$, the change is in reasonable agreement with the small signal gain result $e^{G_0} - 1$. When $I_1^{\mathrm{in}}$ = $2\times10^{14} \mathrm{W/cm^2}$, the output seed power asymptotes to a value lower than the input seed power. The inset shows the power spectrum of electromagnetic waves at the entrance boundary of the seed and indicates substantial BSRS of the higher intensity seed. (b) Despite SRS of the higher intensity seed, the power lost from the pump is in good agreement with the linear theory including pump depletion 
and collisional absorption.}
\label{fig:f5}
\end{figure}

\begin{figure*}
\includegraphics[scale=1.0]{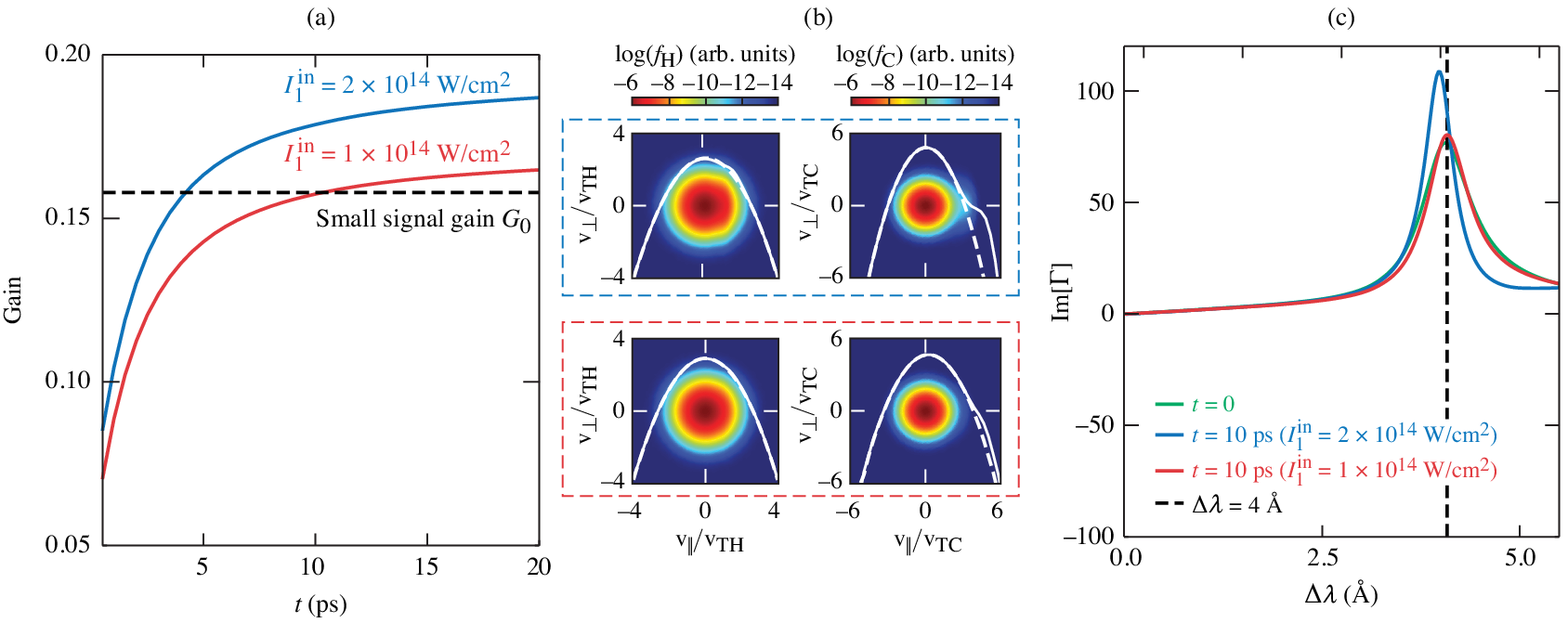}
\caption{\label{fig:wide} (a) Time evolution of the CBET gain for the high pump and seed intensities at 10\% $n_c$. When $I_1^{\mathrm{in}} = 1\times10^{14} \;\mathrm{W/cm^2}$, the gain asymptotes to a value slightly higher than the small signal gain $G_0$. When $I_1^{\mathrm{in}} = 2\times10^{14} \;\mathrm{W/cm^2}$, the gain approaches a value $\sim$$1.3\times$ larger than $G_0$. (b) Trapping in the ion acoustic wave driven by the pump and higher intensity seed flattens the hydrogen and carbon distribution functions in the vicinity of the ion acoustic wave phase velocity. (c) The modified distribution functions at the higher seed intensity redshift the resonant ion acoustic wave frequency and reduce the Landau damping, which narrows and enhances the resonant peak. The enhancement in the peak is consistent with the elevated gain observed in (a). In (b), the distribution functions are averaged over the interaction region at 10 ps. The smooth and dashed lines are the 1D projection of the distribution functions onto the propagation direction of the ion acoustic wave \textbf{k} at $t=0$ and $t=10$ ps, respectively.}
\label{fig:f6}
\end{figure*}

\subsection{PUMP DEPLETION AT 30\% $n_c$}

Figure 4 displays the time evolution of the CBET gain at 30\% $n_c$. For both seed intensities, the gain asymptotes to a value that is consistent with linear theory including pump depletion $G_D$ (Eqs. \ref{eq:tang} and \ref{eq:GD}), but lower than the small signal gain $G_0$. The predominance of pump depletion at 30\% $n_c$ is explained by three factors. Foremost, the density scaling of the kinetic coupling factor $\mathrm{Im}[\Gamma] \propto (n_e/n_c)(1-n_e/n_c)^{-1/2}$ increases the rate of energy transfer and pump depletion at higher densities. Second, the increased rate of ion-ion collisions at higher densities reduces the amount of time that ions stay trapped in an ion-acoustic wave, which inhibits trapping-induced modifications to the ion distribution function. Finally, SRS cannot occur at 30\% $n_c$, because frequency matching is impossible: $\omega_0 < 2\omega_p$ (see Appendix B).

\subsection{RAMAN SCATTERING OF THE SEED AT 20\% $n_c$}

At 20\% $n_c$, BSRS reduces the output power of the higher intensity seed. In an experiment, a reduced output power could be confused with a lower gain. Figure \ref{fig:f5}(a) shows the percent change in the seed power ($P_1^{\mathrm{out}}/P_1^{\mathrm{in}}-1$) as a function of time. The power of the lower intensity seed asymptotes to a value slightly less than that predicted by the small signal gain. The power of the high intensity seed undergoes an initial rise over the first $\sim$2 ps, followed by a rapid drop to a value below the input power. The inset in Fig. \ref{fig:f5}(a) displays the power spectrum of electromagnetic waves at the entrance boundary of the seed over the first 8 ps. At the higher seed intensity, there is a substantially elevated signal at $\omega = \omega_1 - \omega_p$, indicating resonant decay of the seed light into backscattered light and an electron plasma wave, i.e., BSRS. 

The backscattered power into frequencies around $\omega_s \equiv \omega_1 - \omega_p$ is consistent with the Manley-Rowe relations and inverse-Bremsstrahlung (IB) absorption. Integration of the power spectrum around $\omega_s$ indicates that $\sim$15\% of the input seed power ($P_1^{\mathrm{in}}$) was backscattered through the entrance boundary. The total backscattered power, accounting for IB absorption, can be estimated as $P_s \approx 0.15P_1^{\mathrm{in}}\mathrm{exp}(\kappa l)$, where $\kappa = 72.6 \;\mathrm{cm}^{-1}$ is the IB power absorption coefficient modified for a multi-component plasma \cite{yin2012trapping} and $l = 42 \;\mu \mathrm{m}$ is the distance between the top and bottom boundaries of the simulation domain. From the Manley-Rowe relations, the power scattered into electron plasma waves is given by $P_e = (\omega_p/ \omega_{s}) P_{s} \approx 0.18 P_1^{\mathrm{in}}$. Overall, the seed loses $P_s + P_e \approx 0.38 P_1^{\mathrm{in}}$, which agrees with the difference between the linear prediction and VPIC result shown in Fig. \ref{fig:f5}(a).

In this configuration, the seed beam propagates tangential to the density gradient and experiences a near-uniform plasma. This allows SRS to remain phase matched over an extended distance, which in combination with the high density, exacerbates the instability growth. The pump beam propagates parallel to the density gradient, which mitigates SRS by limiting the distance over which it can remain phase matched. As a result, the pump beam loses energy due to CBET but does not undergo SRS. Figure \ref{fig:f5}(c) demonstrates that the power lost from the pump agrees with the linear CBET theory including pump depletion when IB absorption is taken into account ($\sim$9\% of the incident pump power).

\subsection{ION TRAPPING AT 10\% $n_c$}

At 10\% $n_c$, ion trapping in the driven ion acoustic wave reduces the Landau damping and enhances the gain of the higher intensity seed. As shown in Fig. \ref{fig:f6}(a), the gain of the lower intensity seed approaches a value only slightly above the small signal gain $G_0$ due to the finite number of speckles. The gain of the higher intensity seed, on the other hand, approaches a value nearly $1.3\times$ larger than $G_0$. The higher intensity seed drives a larger amplitude ion acoustic wave that traps and accelerates more ions. The trapped ions flatten the distribution functions in the vicinity of the ion acoustic wave phase velocity, $\textrm{v}'_p = 3.7 \textrm{v}_{\mathrm{TC}} = 1.06 \textrm{v}_{\mathrm{TH}}$, where $\textrm{v}_{\mathrm{TC}}$ and $\textrm{v}_{\mathrm{TH}}$ are the carbon and hydrogen thermal velocities [Fig. \ref{fig:f6}(b)]. The flattening reduces the Landau damping and redshifts the resonant ion acoustic wave frequency.\cite{williams2004effects}

Figure \ref{fig:f6}(c) illustrates the effects that the modified distribution functions have on the kinetic coupling factor and ion acoustic wave resonance. The ion acoustic wave driven by the lower intensity seed does not have sufficient amplitude to appreciably modify the distribution functions or the coupling factor. As a result, the resonance peak at $t=10$ ps is identical to that at $t = 0$. The reduction in Landau damping at the higher seed intensity narrows the resonance width and strengthens the response, while the flattening of the distributions redshifts the resonance peak. The elevated response at the drive detuning, $\Delta \lambda = 4.0$ \AA, is consistent with the enhanced gain observed in Fig. \ref{fig:f6}(a). Despite this enhancement, the gain at 10\% $n_c$ is still much lower than that at 30\% $n_c$ due to the density scaling of $\mathrm{Im}[\Gamma]$.

Ion trapping has a larger impact on CBET in lower density and colder plasmas. As illustrated by Fig. \ref{fig:f2}(c), CBET is not energetically significant in the low density region of a direct-drive corona. This suggests that, even at higher intensities, trapping is unimportant. At higher densities, ion-ion collisions are more frequent.\cite{huba20132013} These collisions can either detrap ions or prevent their trapping altogether, thereby inhibiting flattening of the distribution function. At higher temperatures, collisional thermalization of trapped ions produces a smaller relative increase in the ion temperature. SRS and pump depletion, on the other hand, can have a larger impact at higher densities where CBET is more energetically significant: Both the growth rate (see Appendix B) and kinetic coupling factor increase with density.

\section{\label{sec:level1} SUMMARY AND CONCLUSIONS}

Comparisons of 2D collisional PIC simulations with a commonly used reduced, linear model indicate that nonlinear effects do not play a role in CBET for plasma conditions relevant to implosions at the OMEGA laser facility. The PIC simulations were initialized with plasma conditions extracted from a radiation hydrodynamics simulation of an OMEGA implosion. The beam intensities were initialized with average intensities calculated using a 3D ray-based model of CBET. \cite{edgell2017mitigation} The CBET gain predicted by the linear model and the PIC simulations are in excellent agreement. 

For higher intensities, of more relevance to shock or fast ignition, several nonlinear effects were observed to modify CBET, depending on the density. At the highest density, $n_e = 30\% \: n_c$, the strong coupling between the crossing beams resulted in significant pump depletion and gains lower than the linear, small signal gain. At $n_e = 20\% \: n_c$, the seed beam was unstable to BSRS, which reduced the output seed power and apparent energy transfer.  At the lowest density, $n_e = 10\% \: n_c$, ion trapping decreased the Landau damping and enhanced the gain well above the prediction of linear theory.

\begin{acknowledgments}
This material is based upon work supported by the Department of Energy National Nuclear Security Administration under Award Number DE-NA0003856, the University of Rochester, and the New York State Energy Research and Development Authority.
LANL work was performed under the auspices of the U.S. Department of Energy by the Triad National Security, LLC Los Alamos National Laboratory, and was supported by the LANL Office of Experimental Science Inertial Confinement Fusion program. VPIC simulations were run on the LANL Institutional Computing Clusters.

This report was prepared as an account of work sponsored by an agency of the U.S. Government. Neither the U.S. Government nor any agency thereof, nor any of their employees, makes any warranty, express or implied, or assumes any legal liability or responsibility for the accuracy, completeness, or usefulness of any information, apparatus, product, or process disclosed or represents that its use would not infringe privately owned rights. Reference herein to any specific commercial product, process, or service by trade name, trademark, manufacturer, or otherwise does not necessarily constitute or imply its endorsement, recommendation, or favoring by the U.S. Government or any agency thereof. The views and opinions of the authors expressed herein do not necessarily state or reflect those of the U.S. Government or any agency thereof.
\end{acknowledgments}

\section*{Data Availability Statement}

The data that support the findings of this study are available from the corresponding author upon reasonable request.

\appendix
\section{WAVELENGTH DETUNING AND FLOW}

The VPIC simulations presented above used wavelength detuning as a surrogate for flow and did not include a flow gradient. The density and flow in VPIC are constrained by the continuity equation: 
$\partial_t n + \nabla\cdot(n\textbf{u}_f) = 0$. At any particular time, the corona is not in a steady hydrodynamic state, such that $\nabla\cdot(n\textbf{u}_f) \neq 0$. As a consequence, initializing VPIC with a non-uniform density and flow profile from a radiation hydrodynamics simulation would result in unwanted evolution of the density profile. A hydrodynamic equilibrium can be achieved in VPIC by using (1) a non-uniform density and setting $\textbf{u}_f = 0$ or (2) a uniform density and a uniform flow velocity. 

Ultimately, option (1) was chosen for the simulations presented in this paper. Simulations using option (2) predicted substantial levels of BSRS from the pump beam. This does not occur in OMEGA implosions because the density gradient stabilizes BSRS. Thus, the inclusion of the density gradient was critical for faithfully modeling CBET in OMEGA-relevant conditions. Nevertheless, option (2) allows for comparisons of CBET between wavelength-degenerate beams in the presence of flow and wavelength-detuned beams in the absence of flow. These comparisons were used to verify that wavelength detuning is indeed a good surrogate for flow. 

\begin{figure}
\includegraphics[scale=1]{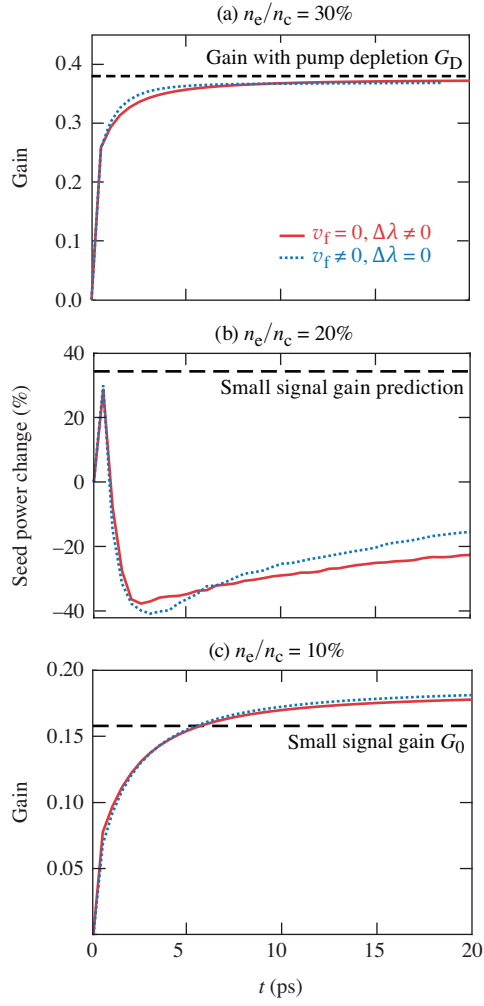}
\caption{Comparison of CBET between wavelength-degenerate beams in the presence of flow and wavelength-detuned beams in the absence of flow. In all cases, the results are nearly identical. The simulations were initialized with a uniform density plasma to avoid unwanted evolution of the density profile in the presence of flow. (a) and (c) Time evolution of CBET gain at 30\% and 10\% $n_c$, respectively. (b) Percentage change of the seed power as a function of time at 20\% $n_c$.}
\label{fig:f7}
\end{figure}

Figure \ref{fig:f7} compares the evolution of CBET when resonant excitation of the ion acoustic wave is achieved through wavelength detuning or flow. In all cases, the pump and seed intensities were set to $I_0^{\mathrm{in}} = I_1^{\mathrm{in}} = 2.0\times10^{14} \,\mathrm{W/cm^2}$, corresponding to the high intensity cases of Sec. \ref{sec:HI}. At both 30\% and 10\% $n_c$, detuning and flow resulted in nearly identical gains over the entire duration of the simulation.  Similarly, at 20\% $n_c$, the two resulted in similar levels of BSRS from the seed, which is illustrated by the loss of seed power. Inspection of the pump power at 20\% $n_c$ (not shown) indicated substantial BSRS from the pump, which was not observed in the inhomogeneous plasma. 

While wavelength detuning can be used to emulate flow for the parameters considered here, this is not always the case. Recent work has shown that CBET between wavelength-shifted beams in a stationary plasma can be substantially suppressed compared to CBET between wavelength-degenerate beams in a flowing plasma.\cite{oudin2021reduction,oudin2022cross}  This suppression occurs when many weakly damped ion acoustic waves originating from many different speckles destructively interfere. However, in the coronal plasmas relevant to OMEGA implosions, the Landau damping rate $\gamma$ of the ion acoustic waves is relatively high: $\gamma$ / $\Omega_A$ $\approx$ 0.1. As a result, the driven ion acoustic wave can only travel a fraction of a speckle width before decaying away. Defining the damping length as $L_{\gamma} \equiv \mathrm{v}'_{A}/\gamma$ and using the speckle width $d_s \approx 4\lambda_0f_\#/\pi $, where $f_\#$ is the f-number, one finds $L_{\gamma}/d_s \lesssim 0.2$ over the entire CBET-active region.

As a final note, initializing VPIC in a hydrodynamic equilibrium ($\partial_t n = 0$) and simulating the pump and probe beams away from their turning points precludes frequency shifts due to the time evolution of the density profile, i.e., the Dewandre shift. \cite{Dewandre} For the interactions considered, these shifts are much smaller than the frequency detuning between the beams (Eq. \ref{eq:lambdaseed}), and thus would not have a substantial effect on the results. Accounting for the shifts in Eq. \ref{eq:KCF} would slightly change the resonant angle at each density.

\section{STIMULATED RAMAN SCATTERING}

An electromagnetic wave propagating through a plasma with frequency $\omega_i$ and wavevector $\mathbf{k}_i$ will scatter from fluctuations in the electron density. The scattered wave will have a frequency $\omega_s = \omega_i \mp \omega_f$ and wavenumber $\mathbf{k}_s = \mathbf{k}_i \mp \mathbf{k}_f$ either down or upshifted by that of the fluctuation ($\omega_f$,$\mathbf{k}_f$). When the fluctuation corresponds to a natural mode of the plasma, the downshifted scattered wave can beat with the initial wave to produce a ponderomotive force that resonantly drives the fluctuation. The resonant drive increases the amplitude of the fluctuation, which, in turn, enhances the rate of scattering. 

The stimulated Raman scattering (SRS) instability refers to a special case of this feedback loop where the natural mode is an electron plasma wave. More specifically,  ($\omega_f,\mathbf{k}_f) = (\omega_e,\mathbf{k}_e)$, where  $1+\chi_e(\omega_e, \mathbf{k}_e) \approx 0$, and ($\omega_s,\mathbf{k}_s) = (\omega_i-\omega_e,\mathbf{k}_i-\mathbf{k}_e)$. These relations indicate that SRS cannot occur when $n_e > n_c/4$. At these densities, $\omega_p > \omega_i/2$, which would result in an evanescent scattered wave with $\omega_s < \omega_p$. For densities lower than $n_c/4$, the dispersion relation for SRS is given by\cite{drake1974parametric,forslund1975theory}
\begin{eqnarray}\label{eq:drake}
\begin{aligned}
\relax[\omega_{s}^{2} - |\mathbf{k}_s|^{2}c^{2}-\omega_{p}^2][1+&\chi_e(\omega_e\mathbf{k}_e)] = \\
 &-\tfrac{1}{2}\chi_e(\omega_e,\mathbf{k}_e)(ck_e|\textbf{a}_i^{\mathrm{in}}|)^2,
\end{aligned}
\end{eqnarray}
where $|\textbf{a}_i^{\mathrm{in}}|=(e/m_ec\omega_i)(2I_i/\epsilon_0\mathrm{v_i})^{1/2}$ is the normalized amplitude of the initial electromagnetic wave. Upon working through some algebra, one can show that the growth is maximized for backscattering:
\begin{eqnarray}\label{eq:SRSgamma}
\gamma_{\mathrm{SRS}} \approx \tfrac{1}{4}ck_e(\omega_e/\omega_s)^{1/2}|\textbf{a}_i^{\mathrm{in}}| - \tfrac{1}{2}(\gamma_e + \gamma_s),
\end{eqnarray}
where $\gamma_e$ and $\gamma_s$ are the amplitude damping rates of the electron plasma and scattered wave, respectively.
Using Eq. \ref{eq:drake} and the parameters corresponding to Fig. \ref{fig:f5}, one finds $\omega_{s} \approx 1.18 \omega_{p}$, $\omega_{e} \approx 1.06 \omega_{p}$, and $\gamma_{\mathrm{SRS}} = 1.1 \times 10^{-3} \omega_{p} = 2.6 \times 10^{12} \;\mathrm{s}^{-1}$, all of which are in agreement with the simulations of the high intensity pump and seed beams at 20\% $n_c$.

\nocite{*}
\bibliography{aipsamp}

\end{document}